\documentstyle[prb,aps,epsfig,twocolumn]{revtex}
\draft
\begin{document}
\twocolumn[\hsize\textwidth\columnwidth\hsize\csname
@twocolumnfalse\endcsname
\title{Methyl group dynamics in a confined glass}
\author{A.J. Moreno$^{1}$, J. Colmenero$^{2,3,4}$, A. Alegr\'{\i}a$^{2,3}$,
C. Alba-Simionesco$^{5}$, G. Dosseh$^{5}$,
D. Morineau$^{5,6}$, B. Frick$^{7}$ 
\vspace{0.3cm}}                     

\address{$^{1}$Laboratoire des Verres, CNRS-UMR 5587, Universit\'{e} de Montpellier II,
B\^{a}timent 13, F-34095 Montpellier (France).\\
$^{2}$Dpto. F\'{\i}sica de Materiales UPV/EHU, Apdo. 1072, E-20080
San Sebasti\'{a}n (Spain).\\
$^{3}$Unidad de F\'{\i}sica de Materiales (CSIC-UPV/EHU), Apdo. 1072, E-20080
San Sebasti\'{a}n (Spain).\\
$^{4}$Donostia International Physics Center, Apdo. 1072, E-20080 San Sebasti\'{a}n (Spain).\\
$^{5}$Laboratoire de Chimie Physique, CNRS-UMR 8000, B\^{a}timent 349,
Universit\'{e} de Paris-Sud, F-91405 Orsay (France).\\
$^{6}${\it Permanent adress:} Groupe Mati\`{e}re Condens\'{e}e et Mat\'{e}riaux, CNRS-UMR 6626,
B\^{a}timent 11A, Universit\'{e} de Rennes I, F-35042 Rennes (France).\\
$^{7}$Institute Laue Langevin, BP 156X, F-38042 Grenoble (France).}

\maketitle
\begin{abstract}
We present a neutron scattering investigation on methyl group dynamics in glassy toluene
confined in mesoporous silicates of different pore sizes. The experimental results have
been analysed in terms of a barrier distribution model, such a distribution following from 
the structural disorder in the glassy state. Confinement results in a strong decreasing of
the average rotational barrier in comparison to the bulk state. We have roughly
separated the distribution for the confined state in a bulk-like and a surface-like contribution,
corresponding to rotors at a distance from the pore wall respectively larger and smaller than
the spatial range of the interactions which
contribute to the rotational potential for the methyl groups. We have estimated a distance of
7 {\AA} as a lower limit of the interaction range,
beyond the typical nearest-neighbour distance between centers-of-mass (4.7 \AA).
\end{abstract}

\vspace{0.8 cm}
]
\narrowtext

\section{Introduction}

Porous materials have attracted great interest in the last years,
since they provide a framework for the study of the physical
propierties of solids under restricted geometry \cite{1}. The dynamics of systems
confined in these materials is expected to differ
considerably from that in the bulk state, first of all because the length of the intermolecular interactions
and correlations is limited by the finite size of the pore if the latter is sufficiently small.
Moreover, the interactions between the pore wall and the confined
molecules -which strongly depend on the porous system- would affect the dynamics of
these molecules in a non-well defined surface layer on the confined system.
Due to their simple dynamics \cite{2}, small molecular rotors as ammonia,
methane, ammine ions or methyl groups
can be considered as adequate dynamical probes
of the effects of confinement by comparing the rotational barriers in 
the bulk and in confined geometry \cite{3}.

A thorough neutron scattering study on methyl groups dynamics in bulk glassy toluene
has been recently presented by some of us \cite{4}. The experimental results were successfully
analysed in terms of a well-established barrier distribution model,
such a distribution resulting from the structural disorder present in the glass.
The average barrier of the distribution was found to be notably higher than that of
the $\beta$-crystalline reference phase,
which shows a short-range structure similar to that of the glassy phase \cite{4,5}.
This fact could indicate that the relevant intermolecular interactions
contributing to the rotational potential of the methyl groups extend
beyond nearest-neighbours.

In order to make a comparison with the bulk state
and to shed new light about this question and about the
interaction lengths controlling the parameters of the distribution,
we have carried out neutron scattering measurements
on glassy toluene confined in mesoporous silicates with different pore sizes.

\section{Theoretical aspects}

The usual single-particle model for methyl group dynamics
in crystalline systems at very low temperatures
($T \approx$ 2 K) is that of a rigid rotor tunnelling through a one-dimensional
rotational potential $V(\Phi)$,
which is restricted to keep the rotational symmetry of the methyl group \cite{2}.
In most of cases only the leading threefold term
of the Fourier expansion is retained, higher order corrections being small:
$V(\Phi)=V_3(1-\cos3\Phi)/2$.
The torsional levels are tunnel-split by the overlapping of the neighbouring single-well wavefunctions,
and the energy splitting of the ground torsional level, or rotational frequency $\omega_{\rm t}$, can be observed
in neutron scattering spectra -for barrier heights $V_3$ below $\approx$ 700 K-
in the $\mu$eV-range as two resolution-width inelastic peaks centered at $\pm\omega_{\rm t}$. The
corresponding incoherent scattering function for rotational tunnelling, normalized to scattering from one
hydrogen is \cite{2}:
\begin{eqnarray}
S^{\rm inc}_{\rm MG}(Q,\omega) = [A(Q)+B(Q)]\delta(\omega)  \nonumber \\
+B(Q)[\delta(\omega+\omega_{\rm t})+\delta(\omega-\omega_{\rm t})]
\end{eqnarray}
with $\hbar Q$, $\hbar\omega$ respectively the momentum and energy transfer of the neutron,
$A(Q)=[1+2j_{0}(Qr)]/3$, $B(Q)=2[1-j_{0}(Qr)]/9$, $j_0(Qr)=\sin(Qr)/(Qr)$
and $r$ the H-H distance in the methyl group.

At temperatures typically above $\sim$ 50 K, a picture of thermally activated classical hopping over the barrier is valid.
The corresponding hopping rate follows an Arrhenius-like temperature dependence
$\Gamma = \Gamma_{\infty}\exp(-E_{\rm A}/kT)$, with $E_{\rm A}$ the classical activation energy, defined as the
difference between the top of the barrier and the ground torsional state. $\Gamma_{\infty}$ is a temperature-independent
preexponential factor.
The incoherent scattering function for the classical hopping
regime is \cite{2}:
\begin{equation}
S^{\rm inc}_{\rm MG}(Q,\omega) = A(Q)\delta(\omega)+3B(Q)L(\omega;\Gamma)
\end{equation}
with $L(\omega;\Gamma)$ a normalized Lorentzian of half-width at half-maximum (HWHM) equal to $\Gamma$.

A series of investigations in structural glasses has evidenced the need of introducing a 
barrier distribution $g(V_3)$ to give account for the different features observed
in the spectra of these systems (rotation-rate-distribution-model, RRDM) \cite{6}.
Such a distribution has its origin in the different local environments for the methyl groups resulting from the
structural disorder in the glassy state. According to this approximation, the spectra for the glassy system are
obtained as a superposition of crystal-like spectra (tunnel-like as (1) at $T \approx$ 2 K,
or classical-like as (2) at high temperature) weighted by $g(V_3)$:
\begin{equation}
S^{\rm inc}_{\rm MG}(Q,\omega) = \int S^{\rm inc}_{\rm MG}(Q,\omega,V_3)dV_3
\end{equation}
Numerical relationships \cite{6} between $V_3$, $\hbar\omega_{\rm t}$ and $E_{\rm A}$ allow the transformation
between the corresponding distributions of these quantities $g(V_3)$, $h(\hbar\omega_{\rm t})$
and $f(E_{\rm A})$. By assuming that the preexponential factor
$\Gamma_{\infty}$ is $V_3$-independent \cite{4,6}, $f(E_{\rm A})$ can also be transformed, through
the Arrhenius law for the temperature dependence of the hopping rate, 
into the corresponding distribution of hopping rates $H(\log\Gamma)$. Thus, the general equation (3)
for the RRDM is reduced to the tunnelling and hopping cases by doing the substitutions
$\delta(\omega\pm\omega_{\rm t})\rightarrow\int h(\pm\hbar\omega_{\rm t})d(\hbar\omega_{\rm t})$ in equation (1)
and $L(\omega,\Gamma)\rightarrow\int H(\log\Gamma)L(\omega,\Gamma)d(\log\Gamma)$ in equation (2).
The consistency of the model is reached when the distributions of tunnelling frequencies
and classical hopping rates, obtained respectively at low and high temperature,
follow from the same barrier distribution $g(V_3)$. 

\section{Experimental details}

Neutron scattering measurements were carried out at the backscattering
spectrometer IN16 of the Institute Laue Langevin (ILL, Grenoble, France). 
A wavelength of 6.27 {\AA} was selected providing an energy resolution of 0.5 $\mu$eV (HWHM).
The energy window available by Doppler shift covered from
-15 to 15 $\mu$eV. The scattering angle covered a range from 11 to 149$^{\rm o}$, resulting
in a $Q$-window from 0.2 to 1.9 \AA$^{-1}$. The instrumental resolution was calibrated
by a vanadium sample, which shows purely elastic scattering.
Raw data were corrected for detector efficiency,
scattering from the sample holder and absorption by means of ILL standard programs.

Spectra were taken on three samples of ring-deuterated toluene confined in mesoporous silicates of
the series MCM-41 and SBA-15, which geometry consists 
of a honeycomb-type lattice of parallel cylindrical pores. They 
were synthesized with pore diameters $D$ = 24, 35 and 47 {\AA}, according 
to an hydrothermal procedure described in Ref. [7]. The porous 
geometry parameters were confirmed by neutron diffraction and nitrogen 
absorption experiments. Deuteration of the silanol groups was achieved by 
chemical H/D exchange with no further chemical treatment of the surface. 
Complete filling of the outgased matrices was achieved with an appropriate 
mass of toluene \cite{8}. 

Flat samples of thickness 2 mm were used to get transmissions above 90 \%, allowing to neglect
multiple scattering effects in the $Q$-range investigated, 1.3 - 1.9 ${\rm \AA}^{-1}$,
where the ratio $B(Q)/A(Q)$ is most favourable in the $Q$-window of IN16.
The cooling rate from room temperature to 2 K achieved in the cryostat
was sufficient to get the glassy state of confined toluene.  
Spectra were also taken for the empty matrixes, showing purely elastic scattering in the
investigated temperature interval.

\section{Results}

Figs. 1,2 show the incoherent scattering functions, normalized to maximum unity,
for confined and bulk glassy toluene -the latter taken from the previuos investigation of Ref. [4]-,
obtained after substraction of all the elastic
contributions due to coherent scattering and to incoherent scattering from 
other atoms different from the methyl group protons.
In this way the comparison between the dynamics in the confined and in the bulk state is not affected
by the different ratios from the methyl group incoherent cross section to the total one. 
From a direct visual inspection, it is evidenced that confinement results
in a strong change in the tunnelling
dynamics at 2 K (Fig. 1), and also at the hopping dynamics at  high temperature (Fig. 2).
As previously reported for the bulk state \cite{4}, the experimental data
for the confined samples at 2 K could be reproduced in terms of a superposition of tunnelling frequencies,
as exposed in Section 2, such a distribution following from an
asymmetric Gamma-distribution of rotational barriers $g(V_3)$:
\begin{equation}
g(V_3)=\frac{(p/{\rm e})^{p}}{V_{3_0}\Gamma(p)}
\left(\frac{V_3}{V_{3_0}}\right)^{p-1} \exp\left[-p\frac{V_3-V_{3_0}}{V_{3_0}}\right]
\end{equation}
with $\Gamma(p)$ the Euler Gamma function, $V_{3_0}$ the average barrier
and $p$ an adimensional shape parameter measuring the asymmetry of the function. The corresponding standard
deviation $\sigma_V$ is obtained as $\sigma_V = V_{3_0}/\sqrt{p}$.
The thick lines in Fig.1 are the RRDM theoretical functions for rotational tunnelling
after convolution with the instrumental resolution.  
The obtained parameters for the three pore diameters investigated are given in Table 1.
The error bars are $\sim$ 5\%.
The resulting distributions are of similar width, but much more asymmetric and with a notably lower
average barrier than that of the bulk. A systematic increase of the average barrier is obtained
with growing pore size, though far from the bulk limit in all cases, suggesting an important contribution
of surface effects to the barrier distribution.

In order to check the consistency of the RRDM for the obtained parameters of $g(V_3)$,
we derived the distribution of classical 
hopping rates through Arrhenius relationship from the distribution of classical activation energies
derived from $g(V_3)$. The corresponding RRDM theoretical function for classical hopping 
was constructed for different values of $\Gamma_{\infty}$,
convoluted with the instrumental resolution and compared with the experimental data.
Thick lines in Fig. 2 correspond to the values $\Gamma_{\infty}=$ 9 and 8 meV for confined
and for bulk glassy toluene \cite{4} respectively.
The error bar for $\Gamma_{\infty}$ can be estimated in the order of $\pm$ 3 and $\pm$ 2 meV
for the confined and the bulk cases respectively. 

\section{Discussion}

The single-particle potential for methyl group rotation in a given system is built up by intra- and
intermolecular interactions extending until a certain radius $r_{\rm c}$. The results above described
show that the distribution of potential barriers for methyl group rotation in glassy toluene is 
strongly modified by confinement. This has to be certainly related with the presence of the pore-wall
and the interactions between the wall and the methyl groups. In a rough approximation we can expect
that the methyl groups which are at a distance from the pore wall $r < r_{\rm c}$ will feel the wall
and their single-particle potential will become modified. On the other hand, methyl groups at
$r > r_{\rm c}$ from the pore wall will in principle behave as bulk-like methyl groups. In this framework,
$g(V_3)$ can be separated into a ``bulk-like'' and a ``surface-like''
contribution, corresponding, respectively, to molecules at a distance of the pore 
surface beyond and below $r_{\rm c}$. Previous results corresponding to dynamics of small rotors
in confined crystalline systems seem to support this kind of separation \cite{3}.
From the cylindrical geometry of the pores, and assuming an homogeneus density
of methyl groups within the pores, it is straightforward to calculate
the fraction of surface-like molecules as $f_{\rm S} = 1-(1-2r_{\rm c}/D)^{2}$.
The corresponding surface-like contribution to the total barrier distribution can be calculated as
$g_{\rm S}(V_3) = [g(V_3)-(1-f_{\rm S})g_{\rm B}(V_3)]/f_{\rm S}$, with $g(V_3)$ and $g_{\rm B}(V_3)$
the distributions for confined and bulk toluene defined by the parameters given in Table 1.
For $r_{\rm c}$ below a certain value, this procedure yields to unphysical negative values of
$g_{\rm S}(V_3)$, so that value can be understood as a lower limit of $r_{\rm c}$.
Unphysical distributions were obtained for $r_{\rm c} <$  5, 6 and 7 ${\rm \AA}$ for $D$ = 24, 35
and 47 ${\rm \AA}$ respectively.
Thereby these values correspond to the maximum possible bulk-like contribution to $g(V_3)$ in each case.
Fig. 3 shows the distributions for one of the pore diameters and for the bulk case,
and also the surface contributions for different values of $r_{\rm c}$, including
some of those with unphysical negative values due to a underestimated value of $r_{\rm c}$
(see inset). As the interaction radius $r_{\rm c}$ should not depend on the pore size and
$r_{\rm c}=$7 {\AA} is also compatible with the $g(V_3)$ obtained for the pore diameters
$D=$ 24 and 35 {\AA} (see Table II), we can take 7 {\AA} as a first estimation for $r_{\rm c}$
in glassy toluene. It is noteworhthy that this value has to be considered only as a lower limit, because
larger values could also be compatible with the different $g(V_3)$ obtained.
Measurements at larger pore sizes would be required to give a more accurate estimation of $r_{\rm c}$.
In any case our results seem to support the idea that the relevant interactions on the methyl group extend beyond
the typical nearest-neighbour distance between centers-of-mass (4.7 \AA) \cite{4,5}, as was suggested
by the comparison of the results in the bulk glassy and $\beta$-crystalline phases \cite{4}.
 
\begin{table}
\begin{center}
\begin{tabular}{cccc}
$D$(\AA) & $V_{3_0}$(K) & $\sigma_{\rm V}$ (K) & $p$ \\
\hline
 24 & 260 & 185 & 2.0  \\ 
 35 & 270 & 190 & 2.0  \\
 47 & 300 & 210 & 2.0  \\
bulk & 420 & 200 & 4.4 \\
\hline
\end{tabular}
\end{center}
\vspace{0.1 cm}
\caption{Parameters of the barrier distribution
for the different pore diameters and for the bulk case.}
\label{tab1}
\end{table}

\begin{table}
\begin{center}
\begin{tabular}{ccccc}
$D$(\AA) & $V^{\rm S}_{3_0}$(K) & $\sigma^{\rm S}_{\rm V}$ (K) & $p$ & $f_{\rm S}$  \\
\hline
 24 & 224 & 155 & 2.1 & 0.83   \\ 
 35 & 187 & 124 & 2.3 & 0.64   \\
 47 & 183 & 119 & 2.4 & 0.51   \\
\hline
\end{tabular}
\end{center}
\vspace{0.1 cm}
\caption{Parameters of the surface-like contribution $g_{\rm S}(V_3)$ calculated by assuming
$r_{\rm c}$ = 7 {\AA}. The fraction of surface-like methyl groups for this value $r_{\rm c}$
is also included.}
\label{tab2}
\end{table}

\begin {figure}
\begin{center}
\resizebox{0.65\columnwidth}{!}{
\includegraphics{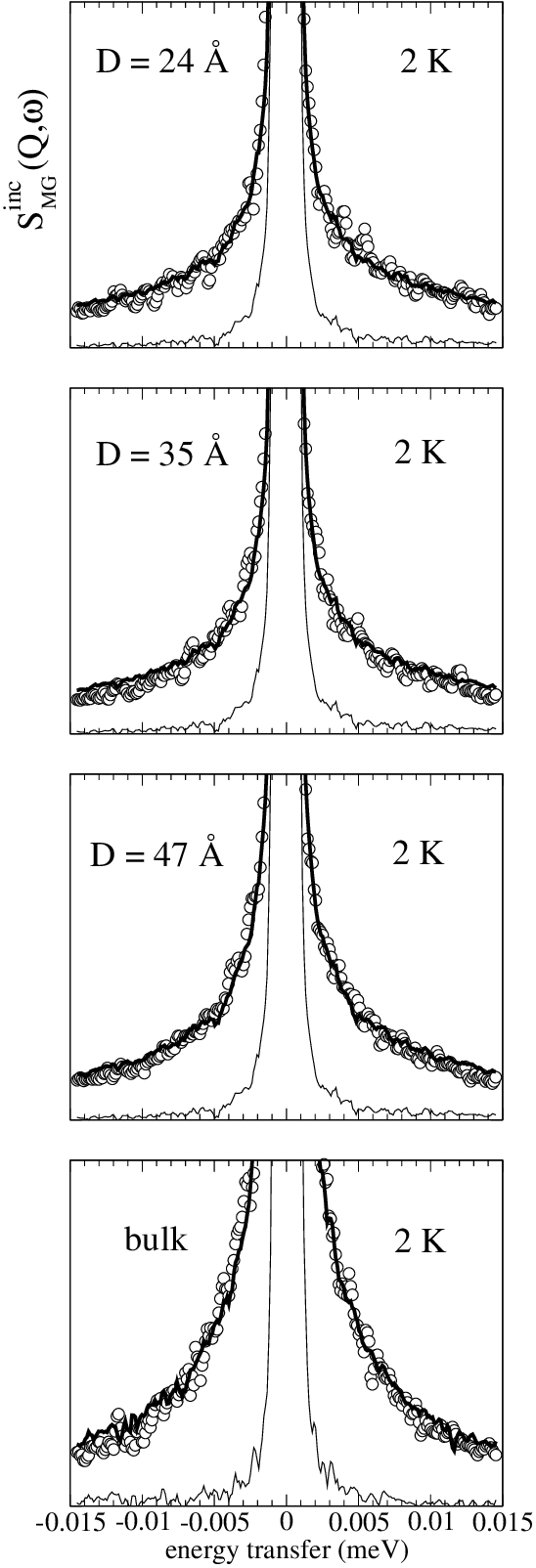}
}
\caption{Incoherent scattering function for methyl group dynamics at 2 K
for confined and glassy bulk toluene. Circles are the experimental data. Thick lines are
the theoretical functions given by the tunnelling version of the RRDM convoluted
with the instrumental resolution (see text).
Thin lines are the instrumental resolution. Amplification scale: 3\% of the maximum.}
\end{center}
\label{figure:1}
\end{figure}

\begin {figure}
\begin{center}
\resizebox{0.65\columnwidth}{!}{
\includegraphics{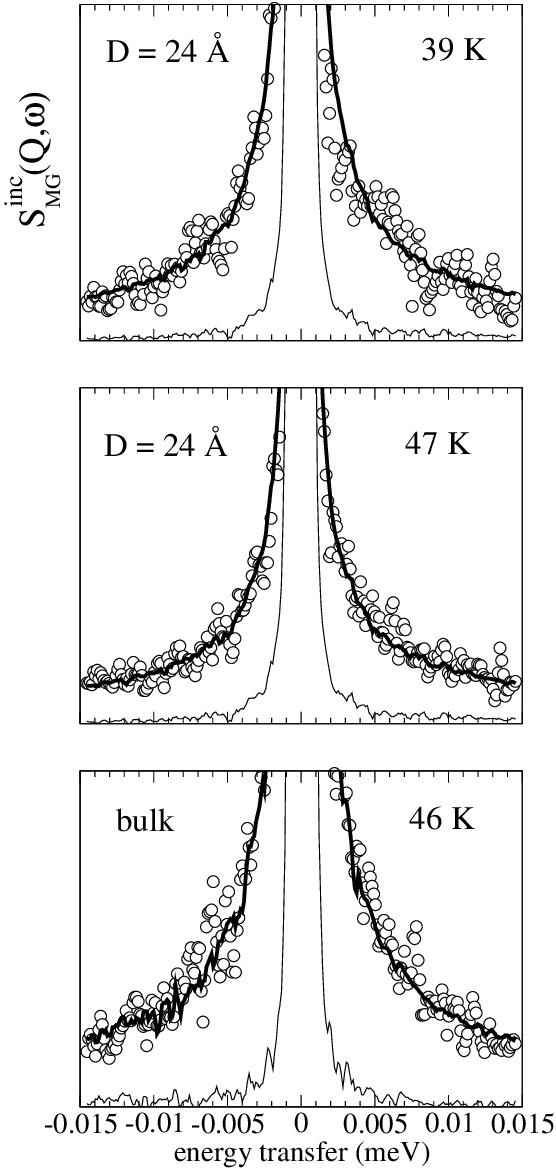}
}
\caption{As Fig. 1 for high temperature. In this case, thick lines are
the theoretical functions given by the classical version of the RRDM.}
\end{center}
\label{figure:2}
\end{figure}

\begin {figure}
\begin{center}
\resizebox{0.9\columnwidth}{!}{
\includegraphics{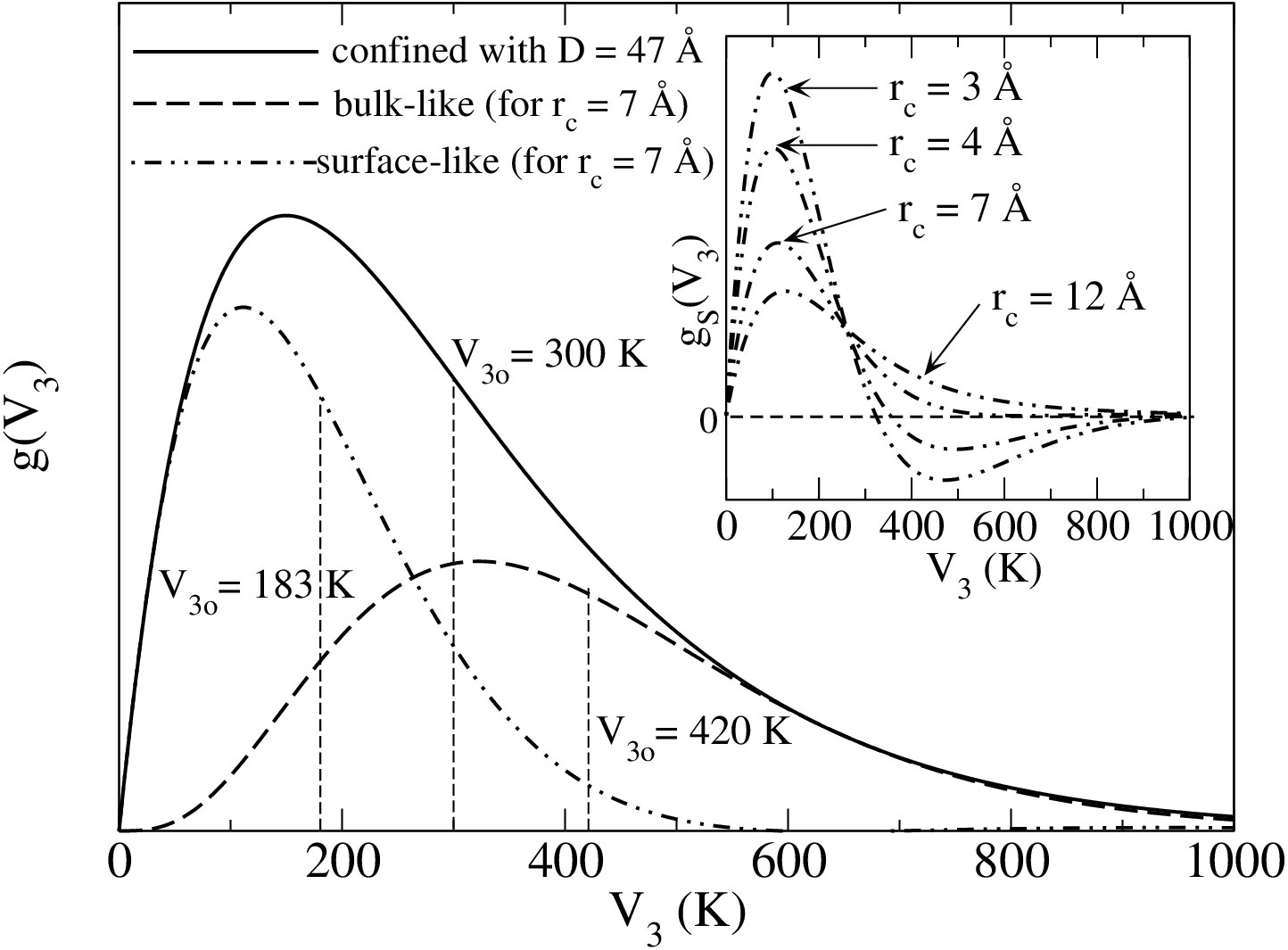}
}
\caption{Barrier distribution $g(V_3)$ for confined glassy toluene with $D$ = 47 {\AA} and its separation
into a bulk-like and a surface-like contribution for $r_{\rm c}$ = 7 {\AA} (see text).
The average barrier is indicated in all cases.
The inset shows the surface-like distribution $g_{\rm S}(V_3)$ for different values of $r_{\rm c}$.
The horizontal dashed line marks the zero value for $g_{\rm S}(V_3)$.}
\end{center}
\label{figure:3}
\end{figure}
%
%

%
\end{document}